\documentclass[final,3p,times,twocolumn]{elsarticle}

\usepackage{amssymb}
\makeatletter
\newcommand\thefontsize[1]{{#1 The current font size is: \f@size pt\par}}
\makeatother
\journal{arXiv}
\begin{document}

\begin{frontmatter}

\title{\textsc{Lux} -- A Laser-Plasma Driven Undulator Beamline}

\author[UHH]{N. Delbos}
\author[UHH]{C. Werle}
\author[UHH]{I. Dornmair}
\author[UHH]{T. Eichner}
\author[UHH]{L. H\"ubner}
\author[UHH]{S. Jalas}
\author[UHH,ELI]{S. W. Jolly}
\author[UHH]{M. Kirchen}
\author[UHH,ELI]{V. Leroux}
\author[UHH]{P. Messner}
\author[UHH]{M. Schnepp}
\author[UHH]{M. Trunk}
\author[UHH,DESY]{P. A. Walker}
\author[UHH,DESY]{P. Winkler}
\author[UHH]{A. R. Maier\corref{cor1}}
\cortext[cor1]{Corresponding author: andreas.maier@cfel.de}

\address[UHH]{Center for Free-Electron Laser Science and Department of Physics, University of Hamburg, Luruper Chaussee 149, 22761 Hamburg, Germany}
\address[ELI]{Institute of Physics of the ASCR, ELI-Beamlines project, Na Slovance 2, 18221 Prague, Czech Republic}
\address[DESY]{Deutsches Elektronen-Synchrotron (DESY), Notkestrasse 85, 22607 Hamburg, Germany}

\begin{abstract}
The \textsc{Lux} beamline is a novel type of laser-plasma accelerator. Building on the joint expertise of the University of Hamburg and \textsc{Desy} the beamline was carefully designed to combine state-of-the-art expertise in laser-plasma acceleration with the latest advances in accelerator technology and beam diagnostics.  \textsc{Lux} introduces a paradigm change moving from single-shot demonstration experiments towards available, stable and controllable accelerator operation. Here, we discuss the general design concepts of \textsc{Lux} and present first critical milestones that have recently been achieved, including the generation of electron beams at the repetition rate of up to 5\,Hz with energies above 600\,MeV and the generation of spontaneous undulator radiation at a wavelength well below 9\,nm.
\end{abstract}

\begin{keyword}
Laser \sep Plasma \sep Wakefield \sep Acceleration
\end{keyword}

\end{frontmatter}


\section{Introduction}
\label{sec:introduction}

Laser-plasma accelerators \cite{cite:esarey} are prominent candidates to drive a future compact, hyper-spectral light source. Providing GeV scale electron energies with bunch lengths of only a few femtoseconds (fs) using centimeter-scale plasma targets, laser-plasma accelerators are ideal to drive a compact undulator and generate single-fs x-ray pulses. An auxiliary beam that is split off the main driver laser just before the plasma target can then drive a variety of optically synchronized secondary sources to provide additional SHG, NIR and THz pulses for pump/probe interaction with the undulator radiation. This approach could form an all-optically driven light source that covers a large spectral range and provides extreme temporal resolution.

However, as the complexity of the experiment after the laser-plasma interaction increases, a controllable, stable and highly available operation of the laser-plasma accelerator becomes crucial. It is essential to make the transition from proof-of-principle demonstration experiments to reliable every-day laser-plasma accelerator operation to enable the complex and demanding experiments downstream of the plasma target.

\begin{figure*}[t]
	\centering
	\includegraphics[width=\textwidth]{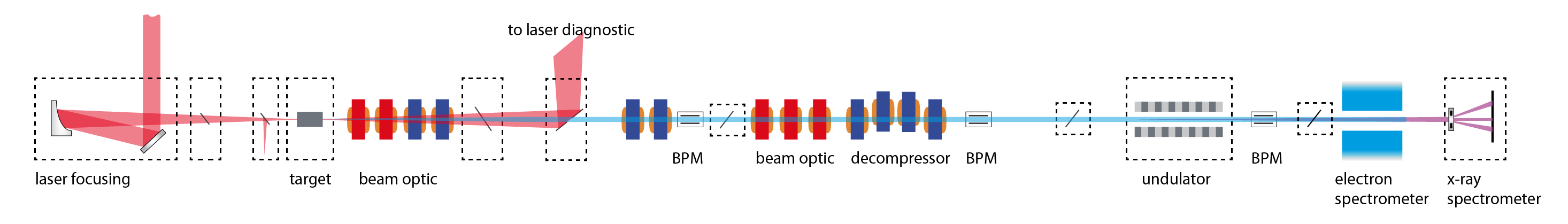}
	\caption{Schematic view of the \textsc{Lux} beamline showing the laser (light red) and electron beam (light blue) propagation after the laser-plasma target. The accelerator has beam optic elements to capture, transport and manipulate the electron beam, including beam position monitors, beam profile monitors, quadrupoles (red) and corrector dipoles (blue). Undulator radiation is detected with transmission grating based spectrometer.}
	\label{fig:LUX-sketch}
\end{figure*}

\textsc{Lux}, as a new facility, was designed from scratch based on three strategic design decisions: (a) Laser quality is understood as the main factor to determine the electron beam quality and great effort has to be invested to generate high-quality laser pulses; (b) high repetition rate electron beams and a robust high rep-rate data acquisition system are the basis for taking data with good statistics; and (c) generating undulator radiation is considered a challenging benchmark for our ability to control the plasma-generated electron beam.

Following these guidelines, the facility is entirely integrated into the accelerator control and data acquisition systems that also operate the user facilities on the campus in Hamburg. The whole vacuum system is designed and built according to machine vacuum standards, and obeys the same specifications as the RF-based accelerators on campus. Our design requires high long-term stability from individual sub-systems, which led to the in-house development of novel accelerator components such as beamline transport mirrors that are carefully characterized for vibration resonances, differential pumping chambers, and a plasma target design based on continuous-flow operation for stable density profiles. The plasma acceleration is driven by the 200\,TW double CPA Ti:sapphire laser system \textsc{Angus}, which has been fully integrated into the controls system, upgraded with enhanced diagnostics, and now features automated feedback loops, enabling daily laser availability with a typical start-up time of less than one hour.

The paper is structured as follows: we outline the scientific goals of \textsc{Lux} in section \ref{sec:goals} and present key milestones that were recently achieved. The technical design of the accelerator is described in section \ref{sec:techDesign} and includes a brief description of the \textsc{Angus} laser performance, the vacuum chamber design, continuous gas flow operation of the plasma target, the \textsc{Lux} undulators and the integration of the facility into the accelerator control- and data acquisition system. Section \ref{sec:program} provides an overview of the research covered by \textsc{Lux}, including the generation of spontaneous undulator and FEL radiation, with an outlook on future developments in section \ref{sec:conclusion}.

\section{Scientific Goals}
\label{sec:goals}

The strategy for the beamline development is based on the three pillars presented above, from which we derive the following scientific goals for \textsc{Lux}: 

\begin{itemize}
\item \textit{Laser development}: The laser has to be available on a daily basis with only minimum start-up time. Stable long-term operation requires a well diagnosed and controlled laser system. The \textsc{Angus} laser is therefore constantly being upgraded with an extensive set of online diagnostics and feedback systems for active stabilization during operation.
\item \textit{High repetition rate laser-plasma acceleration}: To obtain statistically relevant data the rate of electron beam generation from the plasma should only be limited by the available laser repetition rate. This requires a dedicated design for the beamline vacuum chambers and plasma targets, and a gas supply system operating with continuous gas flow and high target pressures, while maintaining UHV conditions in the laser transport beamline \cite{cite:delbos2017}.
\item \textit{Electron phase space manipulation}: To capture, transport and diagnose the electron beam, the beam divergence is of crucial importance. We investigate novel schemes, based on tailoring the longitudinal plasma density profile via the target geometry, to control the beam divergence \cite{cite:dornmair2015} and phase space \cite{cite:brinkmann2017}.
\item \textit{PIC code development and simulations}: PIC codes are an essential tool to understand the processes inside the plasma. We contribute to the development of novel, fast and highly accurate PIC codes \cite{cite:lehe2016a, cite:kirchen2016, cite:lehe2016b, cite:jalas2017} and use them as a support for our experiments.
\item \textit{Spontaneous undulator radiation}: Generation of spontaneous undulator radiation in the soft x-ray range is a challenging benchmark for our ability to capture, transport and control the beam. We generate x-ray pulses in the few nm range to enable first pilot experiments using \textsc{Lux} as a radiation source. 
\item \textit{FEL radiation}: With an FEL as probably the most important application of an all-optical plasma accelerator, we are building a demonstration FEL experiment to show first gain from a plasma-electron beam. We follow the de-compression scheme, that was developed in our group \cite{cite:maier2012}.
\end{itemize}

\section{\textsc{Lux} Technical Design}
\label{sec:techDesign}

Our main goal of building an accelerator based on laser-plasma technology has great influence on the beamline technical design. In general, to achieve a maximum in performance of individual components, we deliberately sacrifice flexibility of the setup. As an example, the whole beamline design is streamlined for high repetition rate, which eventually led us to design a very small target chamber. Simply due to space constraints, the small chamber severely limits our options for different target setups. However, at the same time, the very compact size of the target chamber enables extremely efficient pumping. It is actually a prerequisite for operating the continuous flow target, which took precedence in many design decisions.

\subsection{\textsc{Angus} Laser}

The \textsc{Angus} laser is a double CPA Ti:sapphire laser system, delivering up to 5\,J within 25\,fs on target, i.e. a peak power of 200\,TW, at up to 5 Hz. We measure an energy stability before the compressor well below 1\,\% rms and a beam pointing stability of less than 3 $\mu$rad rms over 500 shots. Using a deformable mirror with a closed-loop for wavefront optimization, we obtain a Strehl ratio of 0.95 calculated from the point spread function retrieved by the wavefront sensor.

The laser is logically separated into different amplification stages with a full set of non-invasive online diagnostics and an automated drift stabilization between consecutive stages. We continuously display key parameters of each amplifier stage for the operators to monitor the system performance. With our control system, we can typically turn on the laser in less than one hour on a daily basis. Continuously monitoring the performance allows us to quickly track down and fix issues during operation.

A pump/probe beam is currently in the process of commissioning. Close to the focussing parabola, we split off $\sim1\,\%$ of the total driver laser energy via a large aperture beamsplitter, which is specially designed for flat transmission of the full laser bandwidth. The transmitted beam is re-compressed to sub 50\,fs using chirped mirrors to compensate for dispersion. The optically synchronized pump/probe beam will be used for a variety of diagnostic experiments and as a driver for secondary sources.

\subsection{\textsc{Lux} Beamline}

\begin{figure}
	\centering
	\includegraphics[width=\linewidth]{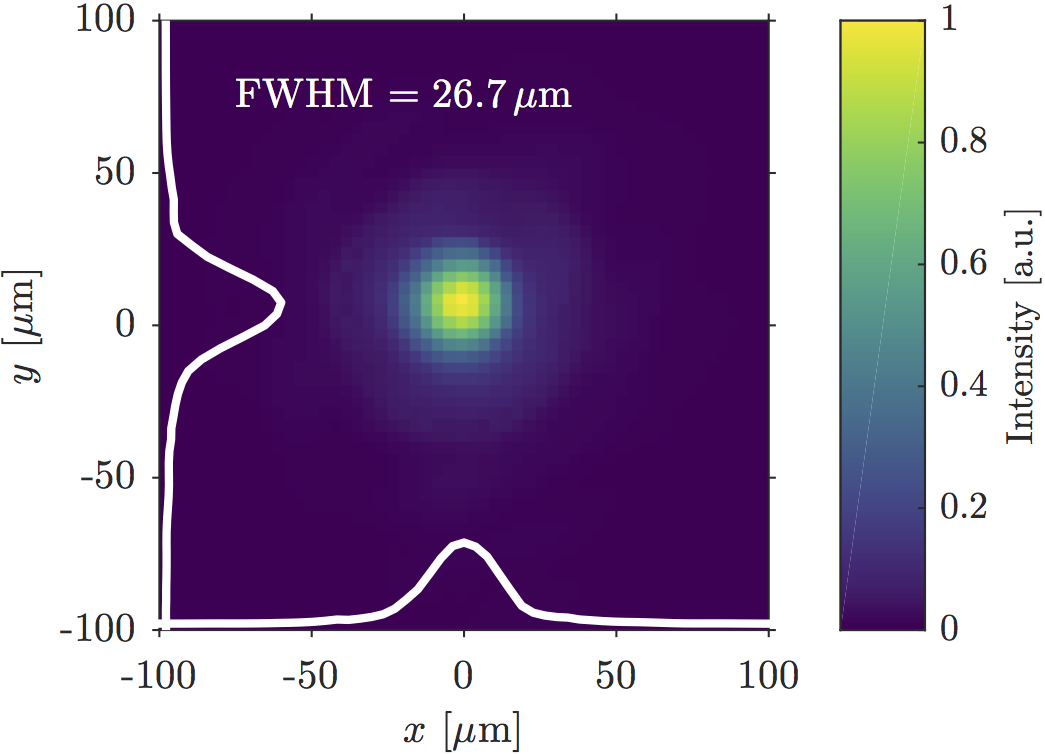}
	\caption{Measured focal spot size of the \textsc{Angus} laser after the 2\,m focal length off-axis parabola. The focal spot FWHM is 26.7\,$\mu$m and close to the diffraction limit.}
	\label{fig:laser:focus}
\end{figure}

The \textsc{Lux} beamline is designed following the \textsc{Desy} machine vacuum standards, which also apply to the conventional UHV accelerator machines on campus. While this requires great technical expenditure by following strict specifications of vacuum technology, a whitelist of suitable materials, and specifications of the residual gas content, it has several benefits: (1) we can connect the \textsc{Angus} laser to \textsc{Regae}, a conventional S-band 5 MeV machine to perform first external injection of a conventional beam into a laser driven plasma \cite{cite:zeitler2013}; (2) the clean UHV environment reduces desorption to a minimum, protects laser components from contamination and increases their lifetime; (3) the speed of pumping the beamline from atmosphere reduces to about half an hour, which is very useful in daily operation; (4) a clean vacuum environment is essential for sensitive x-ray optics, i.e. soft x-ray transmission filters and cooled CCD cameras.

The vacuum specification exclude basically all hydrocarbons contained in, for example, lubricants, rubber seals or plastics and strongly limits the choice of materials. As a result, all motors are physically separated from the vacuum and their motion is coupled to dry-lubricated in-vacuum mechanics. The \textsc{Lux} beamline consists of a series of vacuum chambers, rather than having one large multi-purpose chamber. All \textsc{Lux} chamber are designed and optimized for minimum volume, and each chamber fulfills a specific task. To passively reduce vibrations, all in-vacuum mirror-holders, vacuum chambers and mounts are carefully designed to minimize resonance frequencies below $60$ Hz.

The \textsc{Angus} laser beam is focused with an off-axis parabola of focal length $f = 2025$\,mm. A focal spot size of $26.7\,\mu$m (FWHM) (Fig.\,\ref{fig:laser:focus}) is measured at the pre-target laser diagnostics, alongside the laser wave front and pulse length. Behind the target, the transmitted laser is coupled out for additional online diagnostics. 

\begin{figure}
\includegraphics[width=1.0\columnwidth]{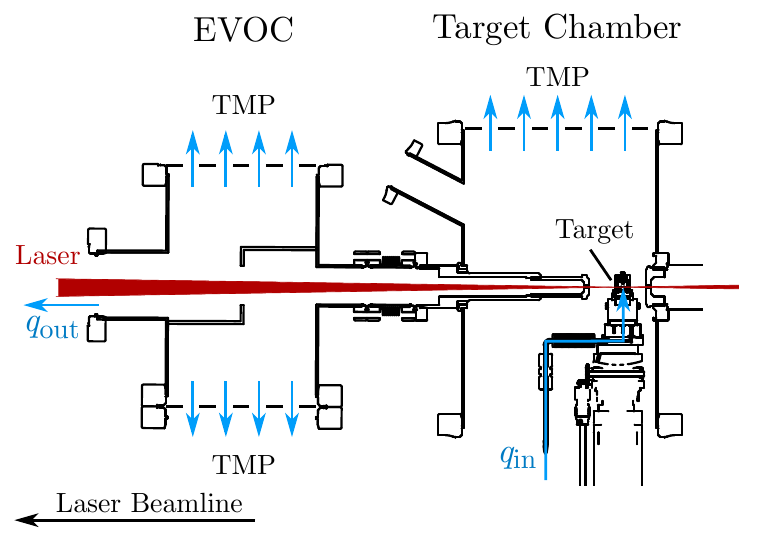}
\caption{Over only 1.5\,m the target and \textsc{Evoc} chambers reduce the pressure level from hundreds of milibars inside the capillary target to the $10^{-6}$\,mbar level in the upstream chambers.}
\label{fig:targetArea}
\end{figure}

To generate plasma electron beams at the full 5\,Hz repetition rate of the laser system requires the plasma target to be supplied with a continuous flow of hydrogen gas, which is in stark contrast to individual gas bursts widely used in laser-plasma accelerators. The gas load of our target is continuously extracted by the dedicated \textsc{Evoc} differential pumping setup connected directly to the target chamber, see Fig.\ \ref{fig:targetArea}. Small apertures strongly restrict the outgoing gas flow from the target chamber. Inside the \textsc{Evoc} chamber a partition wall separates the chamber into two differential pumping volumes on a very small length scale, which then further reduce the gas flow towards the laser beamline. Our pumping concept reduces the pressure by eight orders of magnitude within only $1.5$\,m along the beam axis and allows for pressures on the order of hundreds of millibars inside the target while maintaining $10^{-6}$\,mbar level in the focusing section and upstream laser transport beamline.

\begin{figure}
\includegraphics[width=1.0\columnwidth]{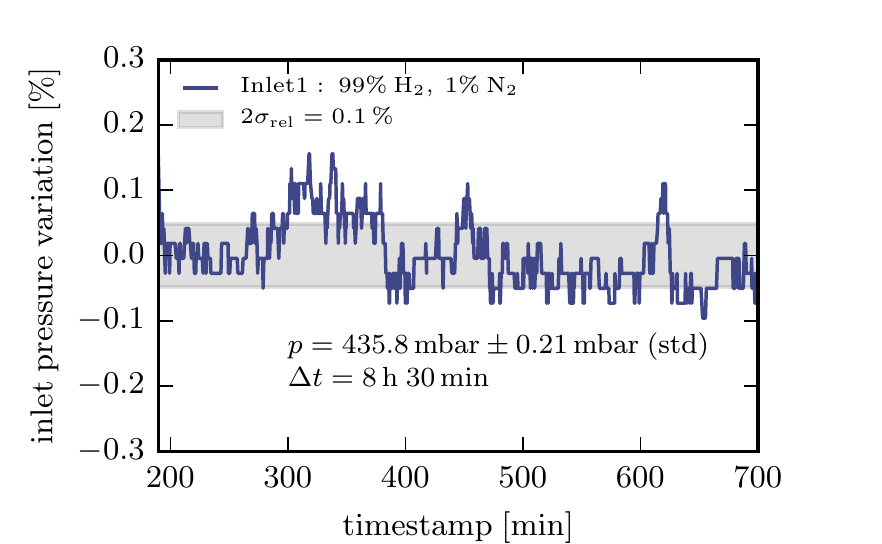}
\caption{Continuous gas flow operation of the \textsc{Lux} target provides long term pressure stability, measured at the inlets of the target, with marginal fluctuations.}
\label{fig:pressureStability}
\end{figure}

Continuous gas flow operation creates a stable, steady state pressure inside the plasma target, which allows us to connect commercially available high-accuracy pressure gauges to the target inlets or the capillary channel. This direct pressure measurement is utilized to monitor the pressure online during accelerator operation. Fig. \ref{fig:pressureStability} shows the long-term pressure and flow stability during continuous flow operation, with fluctuations well below $0.5\,\%$. 

We position the plasma target using a high-precision five-axis stage, based on an x/y/z manipulator for the translation movement, and an in-vacuum goniometer and rotation stage to rotate about the x- and y-axis. All motors are mounted outside of the vacuum system.

After the electron generation we capture the beam using a duplet of electro quadrupoles with individual field gradients of up to 150\,T/m. With this initial configuration of the beam optics we can capture electron beam with energies up to 450\,MeV, which allows us to generate undulator radiation within the water-window. We use two pairs of corrector dipoles to steer the electrons onto the beam axis, which is defined by two cavity BPMs. LYSO and YAG based screens allow us to monitor the electron beam profile.

\subsection{Undulators}
The \textsc{Lux} beamline currently has the \textsc{Beast} undulator installed, with $\lambda_u=5$\,mm and $K=0.3$, to create synchrotron-type (spontaneous) undulator radiation in the soft x-ray range down to a few nm wavelength. First light from the undulator was demonstrated in summer 2017, and it will be used in the future for pilot experiments using \textsc{Lux} as a radiation source. The demonstration FEL experiment \cite{cite:maier2012} uses the \textsc{Frosty} cryogenic undulator, with $\lambda_u=15$\,mm and $K=3$ \cite{cite:bahrdt2015}, which is developed in a close collaboration of our group and the group of J. Bahrdt (HZB). After separation of the electron and photon beam in a permanent magnet dipole spectrometer, we transport the x-rays with a focussing mirror to a transmission-grating spectrometer for single-shot characterization

\subsection{Control System and DAQ}
We use a combination of TINE \cite{cite:tine} and DOOCS \cite{cite:doocs} as the control system and DAQ. Distributed servers control and acquire the data of over 100 diagnostics and operation-critical devices, including CCD cameras, laser spectrometers, laser energy and power sensors, temperature sensors, pressure and flow sensors, motors and actuators, electromagnets, beam position monitors, mass spectrometers, vacuum pumps and vacuum shutters. Interactive panels allow the operators to remotely control the entire experiment in real time.

Almost all diagnostics implement a globally synchronized hardware-triggered eventID for data acquisition at the laser repetition rate. All distributed diagnostic servers then asynchronously publish the generated data over the network together with the unique shot number serving as the time-synchronized identifier. A central data acquisition server processes and stores the shot-to-shot data and serves it over the network for post-processing. The control system enables us to continuously monitor and actively stabilize the accelerator.

\section{Scientfic Programme}
\label{sec:program}
Based on a machine and facility design, which enables plasma acceleration under controlled and closely monitored conditions with high statistics, we follow a carefully coordinated scientific programme.

\subsection{Plasma Target Development}
The electron density along the axis of interaction between laser and plasma strongly determines the laser wakefield acceleration process and the resulting electron beam properties. We systematically study the influence of the longitudinal pressure profile on the electron beam phase space and specifically design the target geometry using OpenFOAM 3D fluid dynamic simulations to tailor application specific longitudinal pressure profiles and electron beams with improved quality \cite{cite:dornmair2015, cite:brinkmann2017}. Here, it is important to separate functional parts inside the plasma target, i.e. electron injection from electron acceleration.

\begin{figure}
\includegraphics[width=1.0\columnwidth]{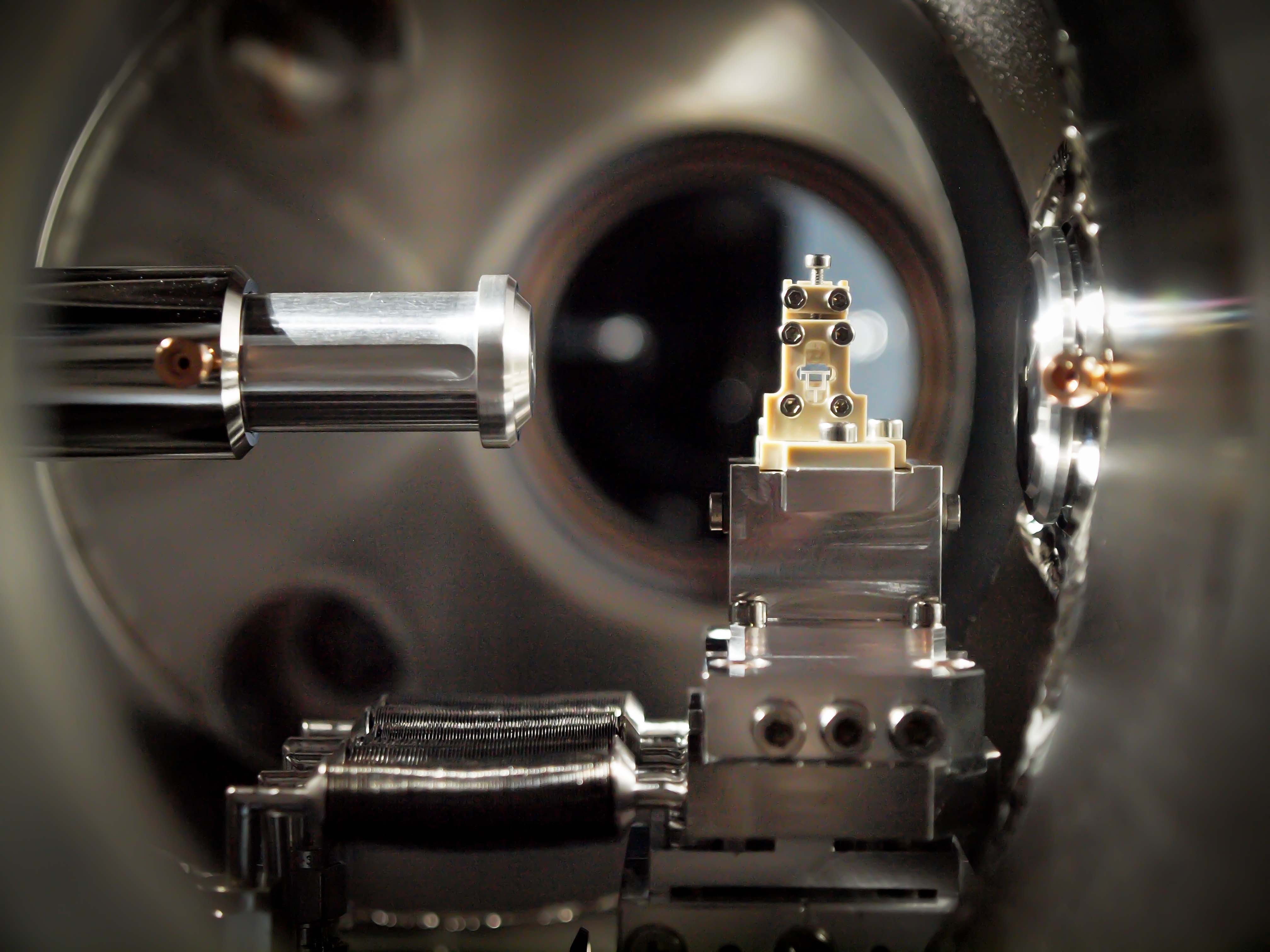}
\caption{The \textsc{Lux} target chamber with a capillary target.}
\label{fig:target}
\end{figure}

\begin{figure}
\includegraphics[width=1.0\columnwidth]{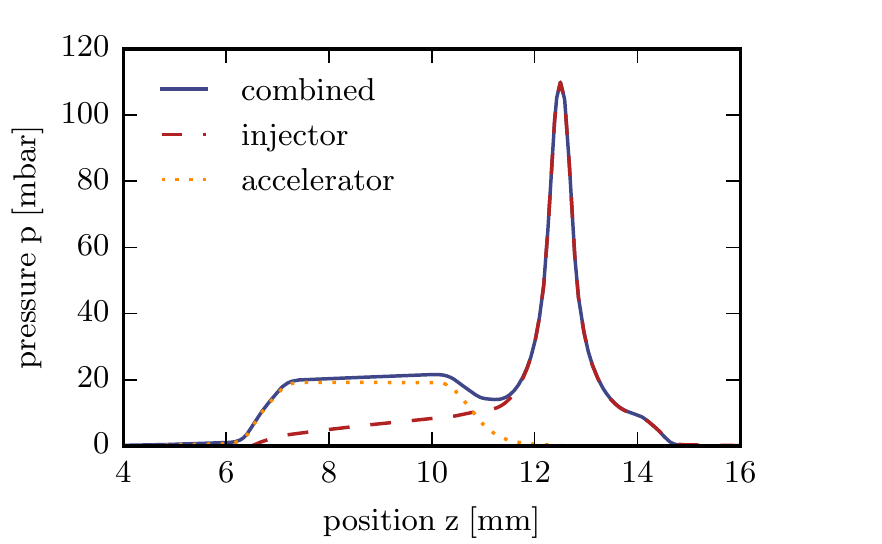}
\caption{Pressure profile (OpenFOAM simulation) with an accelerator section (organe dotted) separated from the injection (dashed-dotted). The geometry of the target strongly suppresses leakage from one stage into the other, allowing for individual control of the profile.}
\label{fig:DDR}
\end{figure}

A basic 4\,mm sapphire capillary target as was used for the demonstration of 5\,Hz electron beams at \textsc{Lux} is shown in figure \ref{fig:target}. Based on this we are following additional target designs.
\textit{Density downramp injection:} The \textsc{Lux} density downramp target design separates the injector stage with a localized pressure peak from the accelerator stage of constant pressure, combined in a single sapphire capillary-type target. Gas leakage from one stage to the other is greatly suppressed, see Fig. \ref{fig:DDR}, such that the pressure in both stages is independently adjustable. In combination with continuous flow operation and a direct pressure measurement, this offers the possibility to independently tune the injection and acceleration pressure during accelerator operation and to scan for optimized electron beam quality. The design also allows to introduce a gas mixture for ionization injection, locally confined to the injector stage of the target. 
\textit{Adiabatic matching:} A tailored density transition from the plasma target to the vacuum will be key to reduce the divergence of the beam, and thus conserve the beam emittance during extraction while minimizing chromatic emittance growth \cite{cite:mehrling2012, cite:floettmann2014, cite:dornmair2015}. We study target designs, which support long up- and downramps at the entrance and exit of a capillary target \cite{cite:delbos2017}.
\textit{Guiding channels:} To reach electron energies on the order of 1\,GeV, reaching well into the water-window using our compact undulator, we use a pulse forming network to guide the laser through the plasma target.

To characterize the longitudinal plasma profile inside the targets, diagnostic techniques are investigated including interferometry, Raman spectroscopy and Stark broadening spectroscopy, which is pioneered by the FLASHForward team on campus \cite{cite:flashforward}. Development of plasma density diagnostics is essential to verify our target designs.

\subsection{Electron Beam Generation}
Using our continuous-flow plasma target, we demonstrated  first laser-plasma driven electron beams at the full 5\,Hz repetition rate of the \textsc{Angus} laser with energies of up to 600\,MeV in 2016. We eventually target for electron beam energies up to 1 GeV using different target designs and guiding the laser inside the plasma. While higher electron beam energies should be possible with the available laser power, our focus is more on electron beam reproducibility, availability and quality, and we tend to trade laser peak power for the benefit of operation with more relaxed parameters. As the experiments downstream from the target chamber become more and more complex it becomes increasingly important to have a highly available and reliable electron source. We could recently show a day-long operation of the plasma accelerator with several 10.000 consecutive shots and high availability of the beam (in preparation).

\subsection{Spontaneous Undulator Radiation}
We want to demonstrate with first pilot experiments the unique properties of an all-optical source, providing tightly synchronized pulses over a larger spectral range. \textit{First soft x-ray performance studies:} With first spontaneous undulator radiation at a few-nm wavelength demonstrated in June 2017, the machine has completed its commissioning phase and we are now performing extensive machine studies. \textsc{Lux} in its current configuration should allow us to generate undulator radiation within the water-window, and, with minor beam optic upgrades, reach above the keV photon energy level. We will correlate the x-ray, electron beam and driver laser properties to gain a sound understanding of the machine, and optimize it for x-ray beam performance, i.e. x-ray photon yield, minimum divergence, and pointing stability. \textit{Synchronization studies:} To demonstrate precise timing we plan to characterize the generated undulator x-ray pulses directly in the time domain, using the THz-based streaking technique, which maps the temporal profile of the x-ray pulse into the kinetic energy of photoelectrons, created by the x-ray pulse focused into a noble gas jet \cite{cite:grguras2012, cite:helml2014}. As short wavelengths minimize slippage effects, temporal characterization of the x-ray pulses could provide a complementary diagnostic for the electron beam.

\subsection{FEL Demonstration}
Our group developed the concept of an FEL demonstration experiment based on longitudinal de-compression to decrease the slice energy spread of a plasma-generated electron beam to a level that allows FEL amplification \cite{cite:maier2012}. This experiment, which is based on the \textsc{Frosty} cryogenic undulator, is extremely demanding, due to the complex beam dynamics of plasma-generated beam \cite{cite:campbell2017}, and will require precise control over the electron beam parameters. The whole development of \textsc{Lux} is an integrated concept with the goal to reach a level of plasma electron beam quality and availability that, in combination with precise control and beam transport, allows to demonstrate first FEL gain.

\subsection{PIC Code Development and Simulations}
We support our experimental activities by numerical studies, trying to verify experimental results, and use the codes to develop new concepts to diagnose and improve the properties of plasma accelerated beams \cite{cite:dornmair2015, cite:dornmair2016ultra, cite:brinkmann2017}. We are contributing to the open-source PIC code FBPIC \cite{cite:lehe2016a} with focus on increasing the speed of the simulations using the Lorentz-boosted frame technique, while eliminating the instabilities and errors from numerical effects \cite{cite:kirchen2016, cite:lehe2016b, cite:jalas2017}.

\section{Conclusion}
\label{sec:conclusion}
After the design and commissioning, \textsc{Lux} is now entering its operation phase following a tight experimental schedule. We already demonstrated the realization of a continuous-flow plasma target and generation of plasma electron beams with 5\,Hz repetition rate. A major milestone was achieved with the demonstration of first x-rays from the \textsc{Beast} undulator at few-nm wavelength. Our actvities heavily use the controls and data acquisition system able to collect data for each single short at few Hz repetition rate. We showed day-long operation of \textsc{Lux} with several 10.000 consecutive shots. This supports our approach of merging our plasma acceleration expertise with state-of-the-art accelerator technology.

\section*{Acknowledgments}
We greatly acknowledge the support and collaboration with ELI Beamlines. We acknowledge funding through DESY, University of Hamburg, BMBF 05K16GU, and H2020 grant No 653782, and the computing time granted on the supercomputer JURECA, project HHH20. Especially, we would like to acknowledge the tremendous support from DESY and University technical groups and workshops.

\end{document}